\documentclass[article,longbibliography,nofootinbib]{revtex4-2}
\usepackage{graphicx} 
\usepackage{xcolor}
\usepackage{tikz}
\usepackage{amsmath}
\usepackage[colorlinks=true, urlcolor=blue, linkcolor=black, citecolor=black]{hyperref}
\usetikzlibrary{arrows.meta, positioning, shapes.geometric,
                 shapes.arrows, calc, backgrounds, fit}

\definecolor{prlblue}{RGB}{31,78,121}
\definecolor{prlred}{RGB}{153,31,35}
\definecolor{prlgray}{RGB}{68,68,68}
\definecolor{prllg}{RGB}{120,120,120}
\definecolor{prlbluetint}{RGB}{234,240,247}
\definecolor{prlredtint}{RGB}{248,234,233}
\begin{document}

\title{\texorpdfstring{A proposal for \\ the safety and controllability requirements that SRM systems should meet}{A proposal for the safety and controllability requirements that SRM systems should meet}}
\author{Eli Waxman}
\thanks{\href{mailto:e.waxman@stardust-initiative.com}{e.waxman@stardust-initiative.com}; $^\dagger$ \href{mailto:a.spector@stardust-initiative.com}{a.spector@stardust-initiative.com}; $^\ddagger$ \href{mailto:y.lederer@stardust-initiative.com}{y.lederer@stardust-initiative.com}}
\author{Amyad Spector,$^\dagger$ Yoav Lederer,$^\ddagger$ Yair Segev}
\author{Tzemah Kislev}
\author{Yanai Yedvab}
\author{Doron Kushnir}
\author{Roby Yahav}

\affiliation{Stardust Labs,  Ness Ziona, Israel; \url{www.stardust-initiative.com}}

\begin{abstract}
    Solar Radiation Modification (SRM) may be the only way to limit global warming in the coming decades, leading to increased interest in the subject and to the expansion of related research \& development (R\&D) activity. Defining the safety and controllability requirements that any SRM system should meet is crucial for directing R\&D activities and enabling governments to make informed decisions on the development and possible implementation of such systems. We present an initial proposal for this set of requirements, which also guides Stardust's R\&D, as a basis for further discussion and consideration. While we focus on SRM systems based on Stratospheric Aerosol Injection (SAI), the proposed principles may be applicable more broadly.
\end{abstract}

\maketitle

\section{Introduction}
\subsection{Why SRM implementation may be required in the near future}

Measurements show a significant increase in global surface temperatures over the last decades, with profound impacts on the Earth's climate system. Since 1850, the so-called pre-industrial times, the global mean surface temperature has risen by more than 1$^\circ$C \citep{IPCCAR6Physical,lee2023ipcc,WMO2025StateClimate}, and the global mean sea level has risen by $\sim0.2$~m \citep{IPCCAR6Physical}. Models predict that, under realistic emission scenarios (SSP2-4.5 - SSP3-7.0 \citep{lee2023ipcc}), the Earth's surface will likely experience an additional warming of 2-3$^\circ$C by the end of this century, and that the global sea level will likely rise by an additional $\sim0.5$~m. 

Observations show a near-linear relationship between global surface temperature increase and atmospheric CO$_2$ concentrations \citep{jarvis2024}. Hence, current mitigation efforts to address global warming focus on reducing the emission of CO$_2$ 
and other greenhouse gases (GHG). Since GHG emissions are currently tied to some of the most fundamental aspects of human activities, including energy production and use, construction, transportation, agriculture, and the production of raw materials \citep{IPCC_WGIII,IEA2021NetZero}, limiting the global average surface temperature rise to less than 1.5°C by the end of the century would require a major global overhaul of infrastructure and processes. Despite growing renewable energy deployment and advances in emissions reductions \citep{forster2025indicators,iea2025weo,unep2025emissions}, such an endeavor is expected to span several decades, even under a well-defined, managed program with worldwide agreement and cooperation.

The widening gap between the goal of limiting global temperature rise and the pace of GHG emission reductions has renewed interest in Solar Radiation Modification (SRM), an approach based on reflecting a small fraction of incoming solar radiation before it reaches Earth's surface \citep{2021NASEM.RS,2023OSTP.SRM,2025RoyalSoc.SRM,2023UNEP.SRMReport,EU_CLP}. Reflecting approximately 1\% of incoming solar radiation, for example, would be sufficient to offset the radiative trapping by the excess GHG accumulated since 1850  
and restore the global mean temperature to pre-industrial levels. The most viable SRM method \citep{2025RoyalSoc.SRM}, which could potentially be implemented within a decade, is Stratospheric Aerosol Injection (SAI) \citep{teller1997global,blackstock2009climate,rasch2008overview,crutzen2006albedo,pope2012stratospheric,weisenstein2015solar,dykema2016improved} - the injection of sub-micron aerosols or aerosol precursors into the stratosphere that would lead to increased sunlight reflection. This method is analogous to the natural process of injecting sulfur-containing species into the stratosphere by volcanoes \citep{McCormick1995_Pinatubo,Soden2002Pinatubo}, which indeed leads to global temperature reductions. If SAI can be safely implemented within a decade, it will enable us to limit global warming and its negative impacts and to "buy the time" required for infrastructure and policy changes that enable emission reductions.

For completeness, we note that carbon dioxide removal (CDR) from the atmosphere is often considered essential for achieving net-zero emissions by counterbalancing hard-to-abate residual emissions, and, if warming exceeds 1.5°C, for sustaining net-negative CO$_2$ emissions to gradually bring temperatures back down \citep{lee2023ipcc}.
However, all suggested removal methods are currently limited in scalability. The current global capacity of engineered carbon dioxide removal from the atmosphere (by novel methods such as direct carbon capture) is $\sim100$~kilotons per year \citep{smith2024state}, and the global capacity of capturing industrially produced carbon dioxide prior to its release into the atmosphere is $\sim40$~megatons per year \citep{smith2024state}. Both are orders of magnitude below the anthropogenic carbon dioxide emission rate, $\sim$~40 gigatons per year \citep{smith2024state}, and are therefore far from being able to significantly modify CO$_2$ concentrations. SRM, by contrast, may be an interim solution to offset warming while GHG emissions are reduced, and should CDR ultimately prove insufficient to fully reverse warming.

\subsection{Goals and scope of the current document}

In view of the growing interest in SRM technology and the possibility that its implementation may be the only feasible way to limit global warming in the coming decades, defining the safety and controllability requirements that any SRM system must meet is both essential and urgent. Such a definition is crucial for directing R\&D activities and enabling governments to make informed decisions on the development and possible implementation of SRM systems. 

Establishing regulatory frameworks for SAI deployment, defining safety and controllability requirements and ensuring they are met by SRM systems, is the responsibility of governments and requires international cooperation. The purpose of this paper is to present an initial proposal for the SAI-based SRM system requirements that reflects our current understanding and guides Stardust's SAI system R\&D efforts\footnote{For more information, visit \url{www.stardust-initiative.com/guiding-principles}.}. We hope it will also serve as a basis for further discussion and consideration.

The paper presents a framework for addressing safety and controllability by identifying risk categories and key risks, and proposing a methodology to address them. It addresses potential risks, uncertainties, and research gaps associated with SAI implementation that were identified in the substantial body of earlier work (e.g., \citep{2021NASEM.RS,2025RoyalSoc.SRM,huynh2024potential,haywood2025wcrp,tilmes2024research,eastham2025key,Tracy2022health,Kravitz2020uncertainty}), as well as "unknown unknowns". It sets quantitative requirements wherever possible, explicitly acknowledges uncertainties, defines approaches to address them, and identifies topics that require further study.
The quantitative requirements, as well as the definitions of some of the criteria, may evolve over time as knowledge and understanding improve through further studies and experience. 

A complete definition of safety requirements should also specify the confidence levels at which the requirements should be met, as well as the methodology to verify that the system satisfies the requirements at those confidence levels. In general, verification would rely on a combination of measurements and modeling that are sufficiently robust and accurate to demonstrate compliance with requirements. While this issue is beyond the scope of the current document, we have taken it into account by setting quantitative requirements that are verifiable with current and plausible near-future measurement and modeling capabilities.

Although the construction of an SAI system can and should prioritize risk reduction and mitigation, risks associated with SAI implementation are inevitable. As for any technological system, meeting safety requirements does not ensure safety at a 100\% confidence level against all risks and under all possible scenarios. Any governmental decision regarding SAI implementation would thus need to be informed by a comparison of the projected risks associated with increased global warming and those related to SAI implementation. We note that safety requirements may limit the magnitude of the temperature reduction achievable by SAI (see, e.g., \citep{Irvine2019Halving}).

A safety-driven approach puts most of the weight on safety rather than on system optimization; considers the safety and controllability of the SRM system as a whole rather than considering only aspects related to the properties of the dispersed particles, and motivates constructing particles that have properties that are derived from requirements and meet strict quality assurance goals for mass production, rather than only focusing on sulfate aerosols analogous with stratospheric injections by volcanoes. In the safety-driven approach, improving efficiency by reducing the total mass of dispersed particles is driven by its impact on risk reduction rather than on cost reduction. 

The safety-driven approach is consistent with earlier work \citep{dykema2016improved,weisenstein2015solar,Vattioni2025,pope2012stratospheric} that argued for exploring solid particle alternatives to sulfate aerosols to reduce risks such as ozone loss and stratospheric heating, and to improve the tunability of radiative forcing. The current paper goes beyond earlier work by identifying key risks of different types, proposing a methodology to address them, and deriving quantitative system requirements from safety and controllability requirements. 

In \S~\ref{sec:guiding} we define the risk categories and discuss the guiding principles for addressing them. Specific risks and the requirements for addressing them are discussed in \S~\ref{sec:requirements}.

\section{Risk categories and guiding principles for addressing them}
\label{sec:guiding}

The risks associated with the implementation of an SAI system are related to three categories of possible adverse or unintended effects: (i) on humans, the biota, and the environment; (ii) on atmospheric chemistry and composition; (iii) on the climate system \citep{2021NASEM.RS,2025RoyalSoc.SRM,huynh2024potential,haywood2025wcrp,tilmes2024research,eastham2025key,Tracy2022health,Kravitz2020uncertainty}\footnote{In category (i) we refer to possible direct impacts due to the presence of dispersed particles; climatic impacts may have additional indirect impacts on humans, the biota, and the environment.}. The safety requirements should correspondingly address the risks of all three categories. It is important to note that these requirements should be addressed by the SAI system as a whole, imposing constraints on all its components: the dispersed particles, the particle dispersal subsystems, the lofting subsystems (e.g., fleet(s) of aircraft carrying dispersal subsystems), and the monitoring subsystems (i.e., remote and/or in-situ sensor networks that monitor both system performance and impact; see below).

For most risks in the first two categories, existing experimental data, combined with laboratory and small-scale field experiments that carry no biological, chemical, environmental, or climatic impacts, are sufficient to substantially constrain uncertainties and determine quantitative safety requirements and to test and assess the margins of compliance of an SAI system with these requirements. Furthermore, for risks associated with possible adverse effects on humans, the biota, and the environment, established requirements and regulations exist that can be largely adopted directly. For risks associated with adverse impacts on atmospheric chemistry and composition, some regulatory precedents exist (e.g., the Montreal Protocol), but a full assessment and, if warranted, a regulatory framework would need to be established.

\begin{figure*}[ht]
\centering
\resizebox{0.5\textwidth}{!}{
\begin{tikzpicture}[
    >=Stealth,
    node distance=0.70cm,
    mainbox/.style={
        rectangle,
        draw=prlgray,
        line width=0.55pt,
        fill=white,
        minimum width=7.6cm,
        minimum height=1.0cm,
        align=center,
        inner sep=7pt,
        font=\normalsize,
    },
    tri/.style={
        fill=#1,
        draw=none,
        regular polygon,
        regular polygon sides=3,
        shape border rotate=180,
        minimum size=6mm,
        inner sep=0pt,
    },
    dbox/.style 2 args={
        draw=#1,
        line width=0.55pt,
        dashed,
        dash pattern=on 5pt off 3pt,
        rounded corners=2pt,
        inner sep=#2,
    },
    sidelbl/.style={
        font=\normalsize\bfseries,
        align=center,
        rotate=90,
    },
]
 
\definecolor{prlbluetint}{RGB}{234,240,247}
\definecolor{prlredtint}{RGB}{248,234,233}
 
\node[mainbox] (inject) {\textbf{Injection of particles}};
\node[tri=prlblue, below=0.22cm of inject] (a1) {};
\node[mainbox, below=0.22cm of a1] (burden) {\textbf{Particles burden profile}};
\node[tri=prlblue, below=0.22cm of burden] (a2) {};
\node[mainbox, below=0.22cm of a2, minimum height=1.25cm] (rf) {%
    \textbf{Radiative forcing}\\[3pt]
    {\small\color{prllg}(SARF\,/\,IRF)}%
};
\begin{scope}[on background layer]
    \node[dbox={prlblue}{12pt}, fill=prlbluetint,
          fit=(inject)(burden)(rf)(a1)(a2)] (bluebox) {};
\end{scope}
\node[sidelbl, text=prlblue] at ([xshift=-8mm]bluebox.west) {%
    RF control\\[2pt]{\small monthly time scale}};
 
\node[tri=prlblue, below=0.85cm of rf, xshift=-2.5mm] (at1) {};
\node[tri=prlred, below=0.85cm of rf, xshift=2.5mm, shape border rotate=0] (at2) {};
 
\node[mainbox, below=1.95cm of rf,
      text width=7.3cm, minimum height=1.5cm] (climate) {%
    \textbf{Climate response}\\[3pt]
    {\small\color{prllg}%
      Temperature\,($\Delta T$),\; precipitation\,($\Delta p$),\; winds,\\[1pt]
      soil moisture,\; extreme events,\; sea level\,($\Delta\mathrm{SL}$)}%
};
\node[tri=prlred, below=0.22cm of climate] (a3) {};
\node[mainbox, below=0.22cm of a3,
      text width=7.3cm, minimum height=1.5cm] (impacts) {%
    \textbf{Impacts}\\[3pt]
    {\small\color{prllg}%
      Agriculture and forestry,\; ecosystems,\; energy\\[1pt]
      production and consumption,\; social effects}%
};
\begin{scope}[on background layer]
    \node[dbox={prlred}{12pt}, fill=prlredtint,
          fit=(climate)(impacts)(a3)] (redbox) {};
\end{scope}
\node[sidelbl, text=prlred] at ([xshift=-8mm]redbox.west) {%
    Climate response\\[2pt]{\small monthly to decadal time scales}};
 
\draw[-{Stealth[length=7pt, width=5pt]}, line width=1.2pt, prlgray]
    ([xshift=10mm]bluebox.north east) -- ([xshift=10mm]redbox.south east)
    node[midway, rotate=-90, above=5pt,
         font=\normalsize\bfseries\color{prlgray}] {Increasing uncertainty};
 
\end{tikzpicture}}
\caption{Cause-effect chain from aerosol injection to climate impacts, following \cite{fuglestvedt2003metrics}. We propose setting the most demanding requirements on the capabilities of the SAI system for producing time- and space-dependent RF profiles: the system should enable the production of a flexible, predictable RF profile and real-time (month-scale) measurements and adaptation of the produced RF. Determining the desired RF profile is beyond the scope of this paper. It should be informed by a sustained community effort to define and prioritize adverse climatic impacts to be minimized, model the relationship between RF and climate impacts, taking into account the spatial and temporal dependence of the climate feedbacks, and draw on experience. Note that many aspects of the climate response depend not only on the RF profile alone, but on state-dependent
feedback mechanisms that evolve with the changing climate system.}
\label{fig:cause_effect_chain}
\end{figure*}
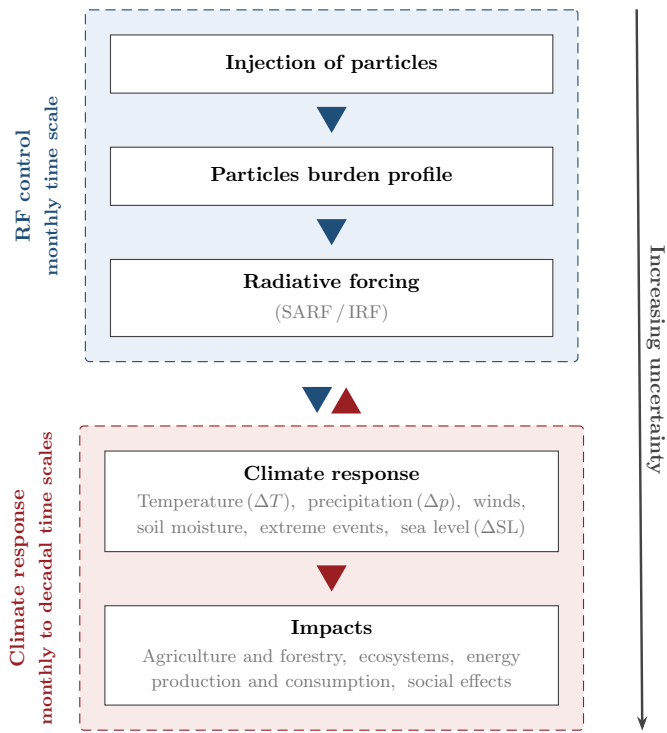

Defining the safety requirements that should be met to address risks in the third category, climatic impact, is more challenging. This is largely due to significant uncertainties in the relationship between the radiative forcing (RF, i.e., the modification by dispersed particles of the radiation flux) that is provided by the SAI system and the climatic impacts (see fig.~\ref{fig:cause_effect_chain}).
\begin{enumerate}
    \item Experimental determination of the climate impact of global SAI implementation cannot be achieved by laboratory or small-scale experiments with negligible climatic impact. Assessing the risks and defining methods to address them will thus require input from theoretical models (see, e.g., \citep{Visioni2023GeoMIP}). While models describing the evolution of the climate system have impressive success in reproducing and accounting for a wide range of observed climate phenomena \citep{IPCCAR6Physical}, the climate system is highly complex, and calculations cannot resolve and account for all relevant processes using first-principles equations. As a result, significant uncertainties and associated risks exist \citep{2025RoyalSoc.SRM}, and will remain in the foreseeable future, in predicting the complex time- and space-dependent response of the climate system to percent-level variations in RF (e.g., resolving regional fluctuations, particularly related to the hydrological cycle, remains challenging). It is important to note that applying SAI in a way that brings temperatures closer to past levels is possible and advantageous, but restoring past conditions cannot be fully achieved due to differences in atmospheric CO$_2$ content \citep{bala2008impact}, suggesting that input from modeling the impact of deviations from past conditions will be needed. 
    \item Experimental determination of the climatic impacts is also challenging. Many climate variables, such as surface temperatures, exhibit large natural long-term (year-scale) variability, which may "mask" SAI effects, particularly during the early stages of deployment when the forced signal is small, thereby limiting their usefulness for 
    assessment of climate impacts (analyses suggest that regional climate anomalies can exceed the SAI-induced signal for years to decades, potentially creating a perception of system failure or unintended effects even when the system is performing as designed \citep{Keys2022perceivedfailure, Lo2016detection}). This underscores the importance of monitoring quantities that can be reliably and accurately measured and on which the SAI system's impact can be reliably and accurately predicted, thus enabling the assessment of the SAI system’s impact during implementation and the adaptation of its strategy (see, e.g., \citep{Wells2024strategies}). Such monitoring will likely need to rely on both existing capabilities \citep{sparc2006assessment,sparc2017data,Kovilakam2023SAGEIII,Loeb2018,Stephens2012,Myhre2025} and newly developed ones.
    \item Finally, even if the climatic impacts of given space- and time-dependent RF profiles were fully known, defining the RF profile that minimizes adverse impacts would depend on the definition and prioritization of adverse impacts that should be reduced and their relative importance. Such a definition and prioritization are yet to be developed and would require a wide community effort (naturally beyond the scope of this paper). Furthermore, many aspects of the climate response depend not on the RF profile alone but on state-dependent feedback mechanisms that evolve with the changing climate system, making any a priori optimization of the forcing profile inherently uncertain.
\end{enumerate}

Due to these limitations, we propose setting the most demanding requirements on the capabilities of the SAI system for producing time- and space-dependent RF profiles: The system should enable production of a flexible and predictable time- and space-dependent RF profile (including gradual ramp-up and termination), and real-time measurements and adaptation of the produced RF (see next paragraph for the definition of "real time"). This set of capabilities, which we refer to as controllability, does not eliminate the climatic impact risks. Rather, it enables choosing an RF profile that minimizes them, and is adjustable in real time to allow modifications informed by modeling and experience. Determining the desired RF profile is beyond the scope of this paper. It should be informed by a sustained community effort to define and prioritize adverse impacts that should be minimized, model the relationship between RF and climate impacts taking into account the spatial and temporal dependence of the climate feedbacks (see, e.g., \citep{Zhao2021LatAlt,Kaur2025HighLat,Crook2011SpatialPO}), and draw on experience. Since the RF's spatial and temporal profiles and their tunability are limited by natural atmospheric properties (see, e.g., \citep{Dai2018Tailoring,macmartin2014spatial}), we require the system to be adjustable in a manner that does not limit the RF's flexibility beyond these natural limitations (see \S~\ref{sub-sec:Phys}).

System requirements are set using measurable quantities that may be spatially and temporally dependent, such as the RF contributed by dispersed particles or the limits on modifying the column density of trace gases in the stratosphere. Since temporal and spatial resolutions are finite, requirements are set on the averaged values of the quantities, and the length and time scales for these averages must be defined. We set the temporal averaging time to 1 month, since this is the time for the fastest response of the atmosphere to RF changes - the time for radiative response and the time for rapid tropospheric response are in the range of weeks. This time scale, 1 month, is therefore also the time required for the SAI system's rapid response. We set the horizontal spatial averaging length scale to $\sim1000$~km (we do not define a vertical averaging length scale since requirements are set with respect to vertically integrated quantities, see below). This is due to the fact that stratospheric meridional transport (across latitude lines) leads to mixing on this scale over $\sim1$~month (zonal mixing, along latitude lines, is much faster), and due to the fact that 1000~km is also the characteristic horizontal length scale for variations in stratospheric flows. We also note that the length scales of climate response to radiative forcing are expected to be larger (due to climate sensitivity, see, e.g., \cite{macmartin2014spatial}). In what follows, we refer to averages over 1~month and 1000~km as "local averages."

Finally, controllability is also key to mitigating the risks of the "unknown unknown" type, i.e., risks that are not identified in advance and may also have adverse impacts in the first two categories, and is important in its own right from governance and public trust perspectives (although we focus here on its role in meeting safety requirements). We note that regulatory frameworks to address unknown risks exist in other areas, like medicine, and should also be developed for SAI. The current document may serve as a basis for such a discussion. 

\section{Risks and safety requirements}
\label{sec:requirements}

\subsection{Safety for humans, the biota, and the environment}
\label{sub-sec:Bio}

Ground-level concentrations of sedimented SAI particles are expected to be much lower than those of already existing aerosols, due to the large injection rate of aerosols into the lower troposphere by natural and anthropogenic sources \citep{colarco2010online}, and due to orders-of-magnitude lower life times of particles in the lower troposphere compared to their stratospheric life time. Nevertheless, any injection of particles that contribute to the lower-troposphere sub-micron particle population should be conducted in accordance with strict safety and environmental regulations. We propose that safety requirements for sub-micron particle populations be derived from, and evaluated in accordance with, established regulatory frameworks, including OECD test guidelines and relevant EU and U.S. regulatory requirements. Compliance with these frameworks requires a comprehensive material budget quantifying the distribution and environmental fate of injected particles and their decay products across relevant environmental compartments (atmosphere, oceans, land, and organisms). 

SAI implementation may affect crop productivity, particularly through the reflection of sunlight that drives photosynthesis. Current studies commonly predict no significant change, and particularly no decrease, in global crop productivity by SAI, due to a combination of factors, including increased diffuse sunlight reaching the ground \citep{Proctor2018crops,Pongratz2012}. No safety requirements are therefore defined here for crop productivity; however, potential impacts on vegetation and agricultural systems should be evaluated.

In light of these considerations, we propose setting the following requirements.
\begin{enumerate}
    \item The particles should be constructed of chemical compounds that have demonstrated compatibility with safety requirements for humans, the biota, and the environment (including limits on impurities \citep{EU_CLP}). We note that some safety tests (e.g. carcinogenicity) may define compounds as safe or unsafe independently of exposure, while others define maximum exposure levels for which the compounds are considered safe - primarily NOAECs (No Observed Adverse Effect Concentrations), defined as the highest tested concentrations that produce no biologically significant adverse effects. Compatibility with the latter type of requirements requires reliable upper limits on expected ground-level exposure.
    \item The dispersed particles’ shape and size distribution should be compatible with these safety requirements.
    The concentration of particles at ground level should be much lower than limits set by regulations, particularly complying with limits on PM2.5 (fine particulate matter with a diameter smaller than 2.5~$\mu$m) and NOAEC (No Observed Adverse Effect Concentrations) criteria for humans and environmental receptors. 
    \item Particle aging (e.g., transformation in the stratosphere or troposphere) and any potential resulting transformation byproducts should be experimentally demonstrated not to affect compliance with the above requirements.
    \item  Particle accumulation at ground level should not affect compliance with the above requirements.
\end{enumerate}
New NOAEC criteria may need to be established (in accordance with accepted methodologies and regulations) if SAI particles contain materials or have a size distribution that extends to small scales for which NOAEC criteria do not exist. Below, we provide a brief explanation of key aspects of implementing the requirements above.

\textbf{Human health:} Standardized toxicity protocols should be applied to evaluate acute and chronic effects through oral, dermal, and inhalation exposure studies to clearly define potential hazards. The implementation should adhere to established guidelines, such as the OECD test guidelines, and follow established methodologies, as defined in the EU-REACH \citep{EU_REACH_2006} and EPA OCSPP \citep{EPA_OPPTS_870}, and could draw additional conclusions from other regulatory framework such as FDA food safety data and pharmaceutical drug delivery research (e.g., 90-day subchronic oral and inhalation toxicity studies, two-year chronic toxicity/carcinogenicity bioassays, two-generation reproductive toxicity assessments, mutagenicity \citep{OECD_Mutagenicity} and carcinogenicity studies using in vitro and in vivo methods \citep{OECD_RepeatedDoseChronic}, reproductive developmental \citep{OECD_ReproductiveDevelopmental} and neurotoxicity evaluation that assesses potential neurological effects \citep{OECD_Neurotoxicity}). Inhalation studies \citep{OECD_AcuteToxicity,OECD_RepeatedDoseChronic} are particularly critical given the atmospheric dispersal pathway. Similarly, dermal irritation, skin sensitization, and eye irritation assessments are necessary to address contact hazards posed by deposited particles \citep{OECD_SkinEyeSensitisation}.
Materials that are found to be hazardous by authoritative bodies (see, e.g., EU-REACH CLP \citep{EU_CLP}, IARC (International Agency for Research on Cancer)) \citep{IARC_Monographs} and "nanoform" materials, with non-negligible sub-0.1$\mu$m particle content (see, e.g., \citep{EU2018_1881}), should be avoided.

To apply quantitative NOAECs within a risk assessment framework intended to protect human health, a benchmark margin of exposure (MOE) is established as the product of appropriate uncertainty or safety factors, which depend on the state of knowledge of the substance under consideration (see, e.g., \citep{EFSA_MOE_2012,ECHA_R8}). The calculated MOE, defined as the NOAEC divided by the estimated human exposure, is then compared against this benchmark to determine whether the exposure is of concern. For ground-level particle concentrations, adequate benchmark MOE values typically fall in the range of 100 to 1000, though values may be lower or higher depending on factors such as the severity of the effect, the steepness of the dose–response relationship, and the completeness of the toxicological dataset. In many cases, this range reflects a 10× uncertainty factor for intraspecies variability and a 10× uncertainty factor for interspecies variability, together ensuring that environmental exposure levels retain a protective safety margin.

\textbf{Biota Health:} Environmental safety relies on aquatic ecotoxicity \citep{OECD_AquaticEcotoxicity} testing using standardized algae, invertebrate, and fish species, complemented by terrestrial studies examining effects on earthworms, plants, birds, pollinators, and soil microbial communities \citep{OECD_TerrestrialEcotoxicity}, establishing environmental no observed effect concentrations (NOECs) that similarly inform safe ground-level concentration limits (e.g. concentrations of concern, COC) through appropriate ecological risk assessment safety margins.

\textbf{Environmental Fate:} Any substance that is introduced into the environment should be assessed for potential bioaccumulation and persistence in the environment. Bioaccumulation assessments extend beyond peak concentration limits to include a comprehensive evaluation of accumulation and elimination pathways, applying established regulatory thresholds from EU-REACH and similar frameworks for aquatic \citep{OECD_AquaticEcotoxicity} and terrestrial organisms \citep{OECD_TerrestrialEcotoxicity}. Particular attention should be given to the potential for tissue partitioning in individual organisms and biomagnification across the food web with available evidence ensuring that clearance mechanisms are sufficient to prevent progressive accumulation under relevant exposure conditions. Evaluating environmental persistence requires demonstrating viable elimination pathways, such as oceanic dissolution, soil degradation, and biological processing, in accordance with standardized persistence criteria \citep{OECD_PersistenceBioaccumulation,EU_REACH_AnnexXIII}.

\subsection{Atmospheric chemistry and composition safety}
\label{sub-sec:Chem}

Injecting particles may impact the chemical composition of the stratosphere through several mechanisms: stoichiometric impacts, arising only from direct (non-catalytic) reactions between the particles and atmospheric constituents and subsequent reactions with their products; catalytic impacts, where the addition or modification of stratospheric aerosol surfaces promotes heterogeneous reactions that impact catalytic cycles, or otherwise impact these cycles by adding or sequestering species which participate in them; and climatic impacts, where the chemical impact originates in the modification of the RF or its distribution. 

The column density of the mass of particles required for providing a 1\% solar flux reflection, $\sim10^{-6}{\rm g/cm^2}$, is relatively small - $10^{-8}$ of the total stratospheric air column density, $1/300$ of the Ozone column density, and comparable to the column density of sulfates and common trace gases such as HCl, NO$_2$, HNO$_3$ and HF \citep{sparc2006assessment,sparc2017data,MERRA}. Therefore, stoichiometric impacts alone, from the release of particle components or reaction products to the stratosphere or the incorporation of stratospheric constituents into the particles, would not directly lead to adverse effects. Rather, the risk of adverse effects is related to the potential enhancement or suppression of catalytic cycles and to the potential promotion of heterogeneous reactions between stratospheric species on particle surfaces. Given the complexity and natural variability of stratospheric kinetics, the ability to predict the impacts of reactive pathways a priori is limited. We therefore propose to address these risks by requiring the particles to be non-reactive in the stratosphere, in the sense that the modification in trace gas concentrations due to their introduction is negligible, and that the modification of ozone concentration due to possible heterogeneous reactions is small compared to its natural variability.

In addition to the promotion of catalytic cycles on bare particle surfaces, catalytic impacts may arise due to modification of the natural stratospheric sulfate surface area by collisions between sulfate aerosols and injected particles or by nucleation of sulfuric acid on such particles \citep{Vattioni2024,Vattioni2023,weisenstein2015solar}. Modification of this surface area is known to impact the ozone column by reducing the concentration of reactive nitrogen oxides, which play a role in terminating halogen cycles \citep{Fahey1993,Prather1992,Hofmann1989,Johnston1992,Mills1993}. We therefore require that the ozone impacts due to such modifications will also be small compared to the natural variability. We note that the natural inter-annual variability of the local sulfate surface area during volcanically quiescent periods is on the order of 10 percent \citep{Liu2012}.

Modification of trace gas concentrations and possible changes in catalytic cycles in the troposphere are not considered a major risk, largely because the lifetime of SAI particles in the lower troposphere is short and their abundance is expected to be small there compared to those of natural or other anthropogenic particles.

In addition to possibly promoting heterogeneous surface reactions, the injected particles may serve as nuclei for cloud formation \citep{Cziczo2019}. This effect may be significant in regions where cloud nuclei are a limiting factor, for example in the upper troposphere, where ice-nucleating particle (INP) abundance is a rate-limiting
factor in cirrus cloud formation \citep{Cirrus2025,Cirrus2025b}. Similarly, in the polar winter stratosphere, polar stratospheric clouds (PSCs) - especially Type 1a clouds, composed of solid Nitric Acid Trihydrate (NAT) - may be constrained by nuclei abundance \citep{PSCs2006,PSCs2023}. Enhanced NAT PSC formation may lead, for example, to a spatial or temporal extension of polar ozone depletion due to enhanced denitrification by sedimenting particles, while modifications to cirrus cloud formation may affect their RF. As the formation processes of upper tropospheric clouds and PSCs are not yet fully understood (as well as the resulting impact of PSC modifications on ozone depletion), conservative safety requirements should be adopted. These may be modified as our understanding improves. 

We therefore propose setting the following requirements.
\begin{enumerate}
    \item Injected particles should not be reactive in the stratosphere. The modification of the locally averaged column densities of stratospheric gases due to direct interactions (i.e., the release of particle components or reaction products to the stratosphere and the incorporation of stratospheric constituents into the particles) should be limited to 0.01 Dobson units (DU; 1~DU~$=2.7\times10^{16}{\rm molecules/cm^2}$), and to 1\% fractional change in the column density of species containing halogens which may act as ozone depletion catalysts (Cl, Br). 0.01~DU corresponds approximately to a few percent of the column density of common, stratospherically important trace gases (HCl, HNO$_3$, NO$_2$) \citep{MERRA}.
    \item The modification of the ozone column density should be smaller than its natural inter-annual variability. The inter-annual variability of the near-global total column ozone (TCO, averaged over all longitudes and latitudes $60^\circ$S to $60^\circ$N) is ${>}3$~DU, and the inter-annual variability during spring at higher latitudes is significantly larger, 20-30~DU ~\citep{Weber2022}. We therefore propose that the SAI modification of the TCO be limited to $\le1$~DU, and that modification of the average column density over higher latitudes in spring be limited to $3$~DU. This limit is well below the ${\sim}10$~DU maximum depletion in TCO observed in the early 1990s~\citep{NOAA2002_OzoneAssessment}. We note that the proposed TCO limit is somewhat lower than the current measurement capability for TCO trends, $0.3 \pm 0.3$\%\,decade$^{-1}$ (${\sim}1 \pm 1$~DU\,decade$^{-1}$) ~\cite{WMO2022}, and below the scale at which current chemistry-climate model ensembles resolve lower-stratospheric ozone trends~\cite{Dietmuller2021,BenitoBarca2025,Abalos2026} and the response to SAI scenarios~\cite{Tilmes2022,Vattioni2025}.
\\    The main mechanisms that may lead to ozone modifications and hence should be limited are:
    \begin{enumerate}
        \item Catalytic ozone depleting reactions, occurring on particle surfaces (heterogeneous reactions), or as a result of chemical interactions of particle constituents.
        \item  Increased surface area of stratospheric sulfates, including mechanisms of wetting of injected particles by, or their coagulation with, background sulfate aerosols, or to particles acting as nuclei for sulfuric acid condensation.
        \item Amplification of PSC impacts, including from the modification of PSC particle sizes and numbers, or extension of their occurrence or coverage to additional times, altitude bands or geographic range. 
        \item Climatic effects of SAI impacting the ozone column, such as changes in stratospheric dynamics, temperature and humidity \citep{Bednarz2023ozone,Vattioni2025} (see also item 4 in \S~\ref{sub-sec:Phys}).
    \end{enumerate}
    \item The modification of upper tropospheric clouds' global mean RF, due to a possible enhancement by SAI particles of INP concentrations \citep{Cziczo2019}, should be limited to 10\% of the instantaneous RF (IRF , see definition in the next section) to enable a predictable RF (see \S~\ref{sub-sec:Phys}). 
    \item Particle aging, as a result of UV irradiation and/or chemical interactions, should be experimentally demonstrated to not affect compliance with the above requirements.
 
\end{enumerate}

\subsection{Climatic effects safety}
\label{sub-sec:Phys}

As discussed in \S~\ref{sec:guiding}, we propose that an SAI system will enable managing the risks associated with possible adverse climatic effects through controllability: the ability to tightly monitor the impacts of the implementation of the SAI system
and to control, monitor, and adapt its operation. The performance requirements are set with reference to the RF following the guidelines explained in \S~\ref{sec:guiding}: the system must enable flexible RF 
(in terms of spatial and temporal dependence) with real-time adjustment capability, that does not limit the RF's flexibility beyond natural limitations. The requirements are set with reference to the direct RF, either the instantaneous RF (IRF), or the stratospherically adjusted RF (SARF) obtained after rapid stratospheric temperature equilibration (which affects infrared emission), rather than to the effective radiative forcing (ERF), which includes feedback from additional rapid processes (like water vapor concentration and cloud modifications). IRF and SARF may be reliably and accurately determined for a given temporal and spatial distribution of particles with given optical properties, while the magnitude and even sign of the impact of the additional feedback processes, which are expected to be small (at the level of a few percent), are uncertain and model-dependent.

In addition to the RF produced by the injected particles, they may impact the climate system by direct heating (or cooling, depending on the particles' optical properties) of the stratosphere. Stratospheric temperature changes affect stratospheric flows and may impact ozone transport and hence chemistry, tropical convection and circulation, and the Earth's large-scale modes of variability such as the quasi-biennial oscillations (QBO) \citep{2025RoyalSoc.SRM,aquila2014modifications,Jones2022, Franke2021QBO}. The assessment of the impacts of stratospheric temperature changes varies significantly between models. We therefore set the safety requirement based on past modifications with observed limited impact. This limit may be updated as our understanding improves.

Any SAI system implementation must undergo validation and verification (V\&V) to reduce climate risk uncertainties and to demonstrate that the system meets the specified requirements, particularly safety requirements. While addressing V\&V processes is beyond the scope of this document, we do include requirements below that should enable verification of the ability to produce a planned pattern of dispersed particles and IRF, through a process of a gradual "ramp-up" during which the amount of dispersed particles is relatively small so as to produce only a negligible RF modification. We address this V\&V aspect due to its unique importance in establishing safety and the significant impact of the associated requirements on the SAI system.

We therefore propose the following requirements.
\begin{enumerate}
    \item The system should allow a gradual "ramp-up" - pilot phase with minimal impact on RF, enabling measurements that demonstrate the ability to produce the planned patterns of dispersed particles and predicted IRF.
    \begin{enumerate}
        \item For a small-scale pilot phase, producing a global IRF corresponding to $\ll0.1\%$ (0.3~W$/{\rm m^2}$) modification of the radiative flux at the top of the atmosphere (a perturbation similar in magnitude to that of the 11-year solar cycle, see e.g. \citep{Coddington2019}), the monitoring sub-system should enable measurements of (locally averaged) stratospheric particle concentrations and size distributions, which enable determining the resulting (locally averaged) IRF (based on known optical particle properties) to $\sim10\%$  accuracy.
        \item For an intermediate scale pilot phase, producing a global IRF corresponding to $\simeq0.1\%$ flux modification, the monitoring sub-system should enable direct measurement of the modification of the global net radiative flux at the top of the atmosphere.
    \end{enumerate}
    \item The system should enable both rapid and gradual shutdown.
    \begin{enumerate}
        \item The SAI particle concentration in the atmosphere should naturally diminish following a deployment halt, leading to a significant reduction (1 e-folding) of the induced RF on a 1-year time scale and to a negligible residual RF level over $\approx5$ years.
       \item The SAI system must support controlled, gradual phase-down over
        timescales of years to a decade to avoid termination shock and to allow flexibility in cessation timescales (see, e.g., \citep{Parker2018}).
    \end{enumerate}        
    \item RF performance.
    \begin{enumerate}
        \item The system should enable a flexible latitude dependence of the IRF and SARF by allowing dispersion at a similar rate per surface area in 20~degree (corresponding to 2000~km) latitude bins. We do not impose constraints on longitudinal dispersion since zonal mixing is likely to prevent effective control of the longitudinal particle distribution. Further analysis accounting for stratospheric flows and mixing along latitude lines is required to substantiate this conclusion.
        \item The system should enable flexible temporal dependence of the IRF and SARF by allowing significant modifications to the RF profile on a month's timescale. 
        \item The locally averaged (over 1~month - 1000~km) IRF and SARF produced by a chosen dispersion pattern should be predictable to better than 10\% (after possible calibration during the ramp-up phase). Note that the uncertainty in estimating the current global GHG RF contribution is approximately 10\% \citep{IPCCAR6Physical}.
        \item The hemispherically averaged IRF and SARF should be predictable and controllable to better than $0.2\,{\rm W/m^2}$. This serves to maintain hemispheric symmetry (see, e.g., \citep{Zhang2024hemispheric}) or, if needed, to allow deviations from symmetry. Meeting this requirement is estimated to be sufficient to enable compensating for the observed Northern Hemisphere–Southern Hemisphere (NH–SH) absorbed solar radiation trend of $+0.34 \pm 0.23$~W\,m$^{-2}$\,decade$^{-1}$ \citep{Loeb2025}, to limit changes in key climate parameters such as the  Inter-tropical Convergence Zone (ITCZ) position and monsoon precipitation index (e.g., for a $0.2\,{\rm W/m^2}$ asymmetry, the estimated shifts are ${\sim}0.2^{\circ}$ and ${\sim}0.7\%$ respectively, if the model-derived sensitivities of \cite{Roose2023} hold), and to provide sufficient control to respond to future hemispherically asymmetric volcanic eruptions.
    \end{enumerate}
    \item The modification of the tropical lower stratospheric temperature (50-100~hPa, 25$^\circ$N-25$^\circ$S) should be limited to 1.5 K.
    This conservative limit is comparable to the stratospheric cooling that accompanied global warming and to the stratospheric heating that accompanied volcanic eruptions such as those of Pinatubo and El Chichón \citep{Free2009}. 
    In addition, meeting this requirement limits the modifications of stratospheric water vapor content and of the Hadley circulation: (i) Observationally constrained estimates indicate stratospheric water vapor increases by $0.31 \pm 0.39$~ppmv~K$^{-1}$ \citep{Nowack2023}, implying that a 1.5~K perturbation would modify stratospheric water vapor by $\lesssim$0.5~ppmv, within 30\% of seasonal variability. (ii) Stratospheric heating increases tropospheric static stability, suppressing deep convection and weakening the Hadley circulation. Idealized simulations that isolate the stratospheric heating pathway from other SAI effects indicate $\sim$0.5\%~K$^{-1}$ weakening of the Northern Hemisphere winter Hadley cell intensity \citep{Cheng2022,Simpson2019}. Under a 1.5~K limit, the implied Hadley cell weakening due to the stratospheric heating pathway alone (independently of any hemispherically asymmetric effects) would be $\lesssim$1\%, well below the $\sim$7\% weakening projected under unmitigated GHG warming by the end-of-the-century \citep{Cheng2022}.
\end{enumerate}

SAI implementation requires monitoring the system performance and its climatic impact. As the Earth's climate system is extensively monitored and studied (see, e.g., \citep{Kovilakam2020GloSSAC, Kovilakam2023SAGEIII}), the measurements and analyses from this activity will serve as a basis for assessing the climate impacts of SAI and for making decisions regarding its adaptation. A full discussion of this topic is outside the scope of this document. Here, we outline the key capabilities of the monitoring subsystem required to determine whether the safety requirements for climatic effects described above are met. As noted in \S~\ref{sec:guiding}, existing experimental data, combined with small-scale laboratory and field experiments that carry no biological, chemical, environmental, or climatic impacts, can and should be used to determine quantitative safety requirements addressing biological, environmental and atmospheric chemistry and composition risks, and to test and assess the margins of compliance of an SAI system with these requirements.

\begin{enumerate}
    \item Particle concentration and size distribution. The system should enable measurements of (locally averaged) stratospheric particle concentration and size distribution, thereby allowing determination of the resulting (locally averaged) IRF to $\sim10\%$ accuracy, for particle column densities exceeding those leading to a global IRF of 0.03~W$/{\rm m^2}$.
    \item RF. (i) The system should enable measurement of a 0.3~W$/{\rm m^2}$ modification of the global net radiative flux at the top of the atmosphere; (ii) The system should enable measurement of a 0.2~W$/{\rm m^2}$ hemispheric asymmetry in the net radiative flux at the top of the atmosphere.
    \item Stratospheric temperature. The monitoring sub-system should enable the measurement of changes of $<1^\circ$C in the locally averaged temperature in the lower stratosphere. We note that the stratospheric temperature response time is weeks, providing relevant information for real-time adaptation of the SAI system. 
    \item Particle tagging. For validating the particle transport model during the gradual "ramp-up" V\&V phase, the ability to determine the origin (time and location) of a particle observed at a given (space and time) point is essential. Tagging particles, in a manner that allows their attribution to specific origin points, would be advantageous for this purpose. Tagging would also be highly useful for governance purposes and small-scale experiments, as discussed above.
\end{enumerate}

\section{Outlook}
\label{sec:outlook}

We have presented a framework for addressing safety and controllability requirements that any SRM system should meet by identifying risk categories and key risks, and proposing a methodology to address them. We have addressed potential risks, uncertainties, and research gaps associated with SAI implementation that were identified in the substantial body of earlier work, as well as "unknown unknowns". Quantitative requirements are set where possible, explicitly acknowledging uncertainties, defining approaches to address them, and identifying topics that require further study. We hope that this will serve as a basis for further discussion and consideration by the community. In subsequent papers, we will describe Stardust's R\&D intended to address the requirements defined in this paper. 

\begin{acknowledgments}
We thank D. Avnir, E. Kandel and S. Magdassi of the Hebrew Univ. of Jerusalem, R. E. Engler and L. A. Holden of Bergeson \& Campbell, P. M. Forster and D. R. Marsh of the Univ. of Leeds, J. I. Katz of Washington Univ. in St. Louis, M. N. Romanias of IMT Nord Europe, B. E. J. Rose of the Univ. at Albany (SUNY), M. R. Schoeberl of Science and Technology Corporation, R. Signorell of ETH Zürich, and J. H. Slade of the Univ. of California, San Diego for useful comments and discussions. 
\end{acknowledgments}

\bibliography{refs-bibfile.bib}

@article{Weber2022,
  author    = {Weber, M. and Arosio, C. and Coldewey-Egbers, M. and Fioletov, V. E. and Frith, S. M. and Wild, J. D. and Tourpali, K. and Burrows, J. P. and Loyola, D.},
  title     = {Global total ozone recovery trends attributed to ozone-depleting substance ({ODS}) changes derived from five merged ozone datasets},
  journal   = {Atmospheric Chemistry and Physics},
  volume    = {22},
  pages     = {6843--6859},
  year      = {2022},
  doi       = {10.5194/acp-22-6843-2022},
  url       = {https://acp.copernicus.org/articles/22/6843/2022/}
}

@article{McCormick1995_Pinatubo,
  author  = {McCormick, M. Patrick and Thomason, Larry W. and Trepte, Charles R.},
  title   = {Atmospheric effects of the Mt Pinatubo eruption},
  journal = {Nature},
  year    = {1995},
  volume  = {373},
  number  = {6513},
  pages   = {399--404},
  doi     = {10.1038/373399a0},
}

@article{Fahey1993,
  author    = {Fahey, D. W. and Kawa, S. R. and Woodbridge, E. L. and Tin, P. and Wilson, J. C. and Jonsson, H. H. and Dye, J. E. and Baumgardner, D. and Borrmann, S. and Toohey, D. W. and Avallone, L. M. and Proffitt, M. H. and Margitan, J. and Loewenstein, M. and Podolske, J. R. and Salawitch, R. J. and Wofsy, S. C. and Ko, M. K. W. and Anderson, D. E. and Schoeber, M. R. and Chan, K. R.},
  title     = {\textit{In situ} measurements constraining the role of sulphate aerosols in mid-latitude ozone depletion},
  journal   = {Nature},
  volume    = {363},
  pages     = {509--514},
  year      = {1993},
  doi       = {10.1038/363509a0},
  url       = {https://www.nature.com/articles/363509a0}
}

@article{Hofmann1989,
  author    = {Hofmann, D. J. and Solomon, S.},
  title     = {Ozone destruction through heterogeneous chemistry following the eruption of {El Chich\'{o}n}},
  journal   = {Journal of Geophysical Research: Atmospheres},
  volume    = {94},
  number    = {D4},
  pages     = {5029--5041},
  year      = {1989},
  doi       = {10.1029/JD094iD04p05029},
  url       = {https://agupubs.onlinelibrary.wiley.com/doi/10.1029/JD094iD04p05029}
}

@article{Johnston1992,
  author    = {Johnston, P. V. and McKenzie, R. L. and Keys, J. G. and Matthews, W. A.},
  title     = {Observations of depleted stratospheric {NO}$_2$ following the {Pinatubo} volcanic eruption},
  journal   = {Geophysical Research Letters},
  volume    = {19},
  number    = {2},
  pages     = {211--213},
  year      = {1992},
  doi       = {10.1029/92GL00043},
  url       = {https://agupubs.onlinelibrary.wiley.com/doi/10.1029/92GL00043}
}

@article{Mills1993,
  author    = {Mills, M. J. and Langford, A. O. and O'Leary, T. J. and Arpag, K. and Miller, H. L. and Proffitt, M. H. and Sanders, R. W. and Solomon, S.},
  title     = {On the relationship between stratospheric aerosols and nitrogen dioxide},
  journal   = {Geophysical Research Letters},
  volume    = {20},
  number    = {12},
  pages     = {1187--1190},
  year      = {1993},
  doi       = {10.1029/93GL01044}
}

@article{Prather1992,
  author    = {Prather, M. J.},
  title     = {Catastrophic loss of stratospheric ozone in dense volcanic clouds},
  journal   = {Journal of Geophysical Research},
  volume    = {97},
  number    = {D9},
  pages     = {10187--10191},
  year      = {1992},
  doi       = {10.1029/92JD00845}
}

@article{Liu2012,
  author  = {Liu, Y. and Zhao, X. and Li, W. and Zhou, X.},
  title   = {Background stratospheric aerosol variations deduced from satellite observations},
  journal = {Journal of Applied Meteorology and Climatology},
  volume  = {51},
  number  = {4},
  pages   = {799--812},
  year    = {2012},
  doi     = {10.1175/JAMC-D-11-0120.1}
}

@article{Vattioni2024,
  author    = {Vattioni, Sandro and Weber, Rahel and Feinberg, Aryeh and Stenke, Andrea and Dykema, John A. and Luo, Beiping and Kelesidis, Georgios A. and Bruun, Christian A. and Sukhodolov, Timofei and Keutsch, Frank N. and Peter, Thomas and Chiodo, Gabriel},
  title     = {A fully coupled solid-particle microphysics scheme for stratospheric aerosol injections within the aerosol--chemistry--climate model {SOCOL-AERv2}},
  journal   = {Geoscientific Model Development},
  volume    = {17},
  pages     = {7767--7793},
  year      = {2024},
  doi       = {10.5194/gmd-17-7767-2024},
  url       = {https://gmd.copernicus.org/articles/17/7767/2024/}
}

@techreport{IEA2021NetZero,
    author      = {{International Energy Agency}},
    title       = {Net Zero by 2050: {A} Roadmap for the Global Energy Sector},
    institution = {International Energy Agency},
    year        = {2021},
    address     = {Paris},
    doi         = {10.1787/c8328405-en},
    url         = {https://www.iea.org/reports/net-zero-by-2050}
}

@techreport{WMO2025StateClimate,
    author      = {{World Meteorological Organization}},
    title       = {State of the Global Climate 2024},
    institution = {World Meteorological Organization},
    year        = {2025},
    number      = {WMO-No. 1368},
    address     = {Geneva},
    isbn        = {978-92-63-11368-5},
    url         = {https://wmo.int/publication-series/state-of-global-climate/state-of-global-climate-2024}
}

@article{MERRA,
    author = {Gelaro, R. and McCarty, W. and Suárez, M.J. and Todling, R. and Molod, A. and Takacs, L. and Randles, C.A.},
    title = {The modern-era retrospective analysis for research and applications, version 2 (MERRA-2)},
    journal = {Journal of climate},
    year = {2017},
    pages = {5419--5454},
    volume    = {30},
    number    = {14},
    doi       = {10.1175/JCLI-D-16-0758.1}
}

@article{PSCs2023,
    author = {Peter Voelger and Peter Dalin},
    title = {Statistical analysis of observations of polar stratospheric clouds with a lidar in Kiruna, northern Sweden},
    journal = {Atmospheric Chemistry and Physics},
    pages ={5551--5565},
    year = {2023},
    volume    = {23},
    number    = {9},
    doi = {10.5194/acp-23-5551-2023}
}

@article{Irvine2019Halving,
    author  = {Irvine, Peter and Emanuel, Kerry and He, Jie and Horowitz, Larry W. and Vecchi, Gabriel and Keith, David},
    title   = {Halving Warming with Idealized Solar Geoengineering Moderates Key Climate Hazards},
    journal = {Nature Climate Change},
    pages   = {295--299},
    year    = {2019},
    volume  = {9},
    number  = {4},
    doi     = {10.1038/s41558-019-0398-8}
}

@article{Dai2018Tailoring,
    author    = {Dai, Zhen and Weisenstein, Debra K. and Keith, David W.},
    title     = {Tailoring Meridional and Seasonal Radiative Forcing by Sulfate Aerosol Solar Geoengineering},
    journal   = {Geophysical Research Letters},
    pages     = {1030--1039},
    year      = {2018},
    volume    = {45},
    number    = {2},
    doi       = {10.1002/2017GL076472}
}

@article{Kaur2025HighLat,
    author  = {Kaur, Harpreet and Bala, Govindasamy and Seshadri, Ashwin K.},
    title   = {Why is the Temperature Response Larger for Radiative Forcing Imposed in High Latitudes than for Forcing Imposed in Low Latitudes?},
    journal = {Climate Dynamics},
    pages   = {211},
    year    = {2025},
    volume  = {63},
    doi     = {10.1007/s00382-025-07659-y}
}

@article{Parker2018,
  author  = {Parker, Andy and Irvine, Peter J.},
  title   = {The Risk of Termination Shock From Solar Geoengineering},
  journal = {Earth's Future},
  volume  = {6},
  number  = {3},
  pages   = {456--467},
  year    = {2018},
  doi     = {10.1002/2017EF000735}
}

@article{forster2025indicators,
    author  = {Forster, Piers M. and Smith, Chris and Walsh, Tristram and Lamb, William F. and Lamboll, Robin and Cassou, Christophe and Hauser, Mathias and Hausfather, Zeke and Lee, June-Yi and Palmer, Matthew D. and von Schuckmann, Karina and Slangen, Aim{\'e}e B. A. and Szopa, Sophie and Trewin, Blair and Yun, Jeongeun and others},
    title   = {Indicators of Global Climate Change 2024: Annual Update of Key Indicators of the State of the Climate System and Human Influence},
    journal = {Earth System Science Data},
    pages   = {2641--2680},
    year    = {2025},
    volume  = {17},
    doi     = {10.5194/essd-17-2641-2025}
}

@article{jarvis2024,
    author  = {Jarvis, Andrew J. and Forster, Piers M.},
    title   = {Estimated Human-Induced Warming from a Linear Temperature and Atmospheric {CO\textsubscript{2}} Relationship},
    journal = {Nature Geoscience},
    volume  = {17},
    pages   = {1222--1224},
    year    = {2024},
    doi     = {10.1038/s41561-024-01580-5}
}

@article{Crook2011SpatialPO,
  title={Spatial Patterns of Modeled Climate Feedback and Contributions to Temperature Response and Polar Amplification},
  author={Crook, Julia A. and Forster, Piers M. and Stuber, Nicola},
  journal={Journal of Climate},
  year={2011},
  volume={24},
  number={14},
  pages={3575--3592},
  doi={10.1175/2011JCLI3863.1}
}

@techreport{unep2025emissions,
    author      = {{United Nations Environment Programme}},
    title       = {Emissions Gap Report 2025: Off Target -- Continued Collective Inaction Puts Global Temperature Goal at Risk},
    institution = {United Nations Environment Programme (UNEP)},
    year        = {2025},
    address     = {Nairobi}
}

@techreport{iea2025weo,
    author      = {{International Energy Agency}},
    title       = {World Energy Outlook 2025},
    institution = {International Energy Agency (IEA)},
    year        = {2025},
    address     = {Paris},
    url         = {https://www.iea.org/reports/world-energy-outlook-2025},
    note        = {Licence: CC BY 4.0}
}

@techreport{lee2023ipcc,
    author  = {Lee, Hoesung and Romero, Jos{\'e}},
    title   = {{IPCC, 2023: Climate Change 2023: Synthesis Report. Summary for Policymakers. Contribution of Working Groups I, II and III to the Sixth Assessment Report of the Intergovernmental Panel on Climate Change}},
    institution = {Intergovernmental Panel on Climate Change (IPCC)},
    address = {Geneva, Switzerland},
    year    = {2023},
    note    = {Core Writing Team, H. Lee and J. Romero (eds.)}
}

@article{Zhao2021LatAlt,
    author    = {Zhao, Minghua and Cao, Long and Bala, Govindasamy and Duan, Lei},
    title     = {Climate Response to Latitudinal and Altitudinal Distribution of Stratospheric Sulfate Aerosols},
    journal   = {Journal of Geophysical Research: Atmospheres},
    pages     = {e2021JD035379},
    year      = {2021},
    volume    = {126},
    number    = {24},
    doi       = {10.1029/2021JD035379}
}

@article{Soden2002Pinatubo,
    author    = {Soden, Brian J. and Wetherald, Richard T. and Stenchikov, Georgiy L. and Robock, Alan},
    title     = {Global Cooling After the Eruption of {Mount Pinatubo}: A Test of Climate Feedback by Water Vapor},
    journal   = {Science},
    pages     = {727--730},
    year      = {2002},
    volume    = {296},
    number    = {5568},
    doi       = {10.1126/science.296.5568.727}
}

@article{Cirrus2025b,
    author = {David L. Mitchell and Anne Garnier},
    title = {Advances in CALIPSO (IIR) cirrus cloud property retrievals–Part 2: Global estimates of the fraction of cirrus clouds affected by homogeneous ice nucleation},
    journal = {Atmospheric Chemistry and Physics} ,
    pages ={14099--14129},
    year = {2025},
    volume    = {25},
    number    = {20},
    doi = {10.5194/egusphere-2024-3814}
}

@article{Cirrus2025,
    author = {Lin, L. and Liu, X. and Zhao, X. and Shan, Y. and Ke, Z. and Lyu, K. and Bowman, K. P.},
    title = {Ice nucleation by volcanic ash greatly alters cirrus cloud properties},
    journal = {Science Advances},
    year = {2025},
    volume    = {11},
    number    = {19},
    doi = {10.1126/sciadv.ads05}
}

@article{PSCs2006,
    author = {Xihong Wang and Diane V. Michelangeli},
    title = {A review of polar stratospheric cloud formation},
    journal = {China Particuology},
    pages ={261--271},
    year = {2006},
    volume    = {4},
    number    = {6},
    doi = {10.1142/S1672251506000534}
}

@article{Cziczo2019,
  author    = {Cziczo, Daniel J. and Wolf, Martin J. and Gasparini, Bla{\v{z}} and M{\"u}nch, Steffen and Lohmann, Ulrike},
  title     = {Unanticipated Side Effects of Stratospheric Albedo Modification Proposals Due to Aerosol Composition and Phase},
  journal   = {Scientific Reports},
  volume    = {9},
  pages     = {18825},
  year      = {2019},
  doi       = {10.1038/s41598-019-53595-3},
  url       = {https://www.nature.com/articles/s41598-019-53595-3}
}

@article{Vattioni2023,
  author    = {Vattioni, Sandro and Luo, Beiping and Feinberg, Aryeh and Stenke, Andrea and Vockenhuber, Christof and Weber, Rahel and Dykema, John A. and Krieger, Ulrich K. and Ammann, Markus and Keutsch, Frank N. and Peter, Thomas and Chiodo, Gabriel},
  title     = {Chemical Impact of Stratospheric Alumina Particle Injection for Solar Radiation Modification and Related Uncertainties},
  journal   = {Geophysical Research Letters},
  volume    = {50},
  number    = {24},
  pages     = {e2023GL105889},
  year      = {2023},
  doi       = {10.1029/2023GL105889}
}

@techreport{OECD_AcuteToxicity,
  author       = {{OECD}},
  title        = {OECD Guidelines for the Testing of Chemicals, Section 4: Health Effects},
  institution  = {Organisation for Economic Co-operation and Development},
  year         = {2002--2022},
  address      = {Paris},
  note         = {Acute toxicity guidelines referenced: 
                  TG 402 -- Acute Dermal Toxicity (2017);
                  TG 403 -- Acute Inhalation Toxicity (2009);
                  TG 423 -- Acute Oral Toxicity, Acute Toxic Class Method (2002);
                  TG 425 -- Acute Oral Toxicity, Up-and-Down Procedure (2022);
                  TG 436 -- Acute Inhalation Toxicity, Acute Toxic Class Method (2009)}
}

@techreport{OECD_RepeatedDoseChronic,
  author       = {{OECD}},
  title        = {OECD Guidelines for the Testing of Chemicals, Section 4: Health Effects},
  institution  = {Organisation for Economic Co-operation and Development},
  year         = {2009--2018},
  address      = {Paris},
  note         = {Repeated dose and chronic toxicity guidelines referenced:
                  TG 407 -- Repeated Dose 28-Day Oral Toxicity Study in Rodents;
                  TG 408 -- Repeated Dose 90-Day Oral Toxicity Study in Rodents;
                  TG 412 -- Subacute Inhalation Toxicity: 28-Day Study;
                  TG 413 -- Subchronic Inhalation Toxicity: 90-Day Study;
                  TG 451 -- Carcinogenicity Studies;
                  TG 452 -- Chronic Toxicity Studies;
                  TG 453 -- Combined Chronic Toxicity/Carcinogenicity Studies}
}

@techreport{OECD_SkinEyeSensitisation,
  author       = {{OECD}},
  title        = {OECD Guidelines for the Testing of Chemicals, Section 4: Health Effects},
  institution  = {Organisation for Economic Co-operation and Development},
  year         = {2004--2024},
  address      = {Paris},
  note         = {Skin, eye, and sensitisation guidelines referenced:
                  TG 430 -- In Vitro Skin Corrosion: Transcutaneous Electrical Resistance Test Method;
                  TG 431 -- In Vitro Skin Corrosion: Reconstructed Human Epidermis Test Method;
                  TG 435 -- In Vitro Membrane Barrier Test Method for Skin Corrosion;
                  TG 439 -- In Vitro Skin Irritation: Reconstructed Human Epidermis Test Method;
                  TG 442C -- In Chemico Skin Sensitisation: Direct Peptide Reactivity Assay;
                  TG 442D -- In Vitro Skin Sensitisation: ARE-Nrf2 Luciferase Test Method;
                  TG 442E -- In Vitro Skin Sensitisation: Human Cell Line Activation Test;
                  TG 467 -- Defined Approaches for Serious Eye Damage and Eye Irritation;
                  TG 492 -- Reconstructed Human Cornea-like Epithelium (RhCE) Test Method;
                  TG 496 -- In Vitro Macromolecular Test Method for Identifying Chemicals Inducing Serious Eye Damage and Not Requiring Classification for Eye Irritation or Serious Eye Damage}
}

@techreport{OECD_Mutagenicity,
  author       = {{OECD}},
  title        = {OECD Guidelines for the Testing of Chemicals, Section 4: Health Effects},
  institution  = {Organisation for Economic Co-operation and Development},
  year         = {1997--2023},
  address      = {Paris},
  note         = {Mutagenicity guidelines referenced:
                  TG 471 -- Bacterial Reverse Mutation Test (Ames Test);
                  TG 474 -- Mammalian Erythrocyte Micronucleus Test (in vivo, if in vitro positive);
                  TG 487 -- In Vitro Mammalian Cell Micronucleus Test;
                  TG 489 -- In Vivo Mammalian Alkaline Comet Assay (if in vitro positive)}
}

@techreport{OECD_ReproductiveDevelopmental,
  author       = {{OECD}},
  title        = {OECD Guidelines for the Testing of Chemicals, Section 4: Health Effects},
  institution  = {Organisation for Economic Co-operation and Development},
  year         = {2018--2022},
  address      = {Paris},
  note         = {Reproductive and developmental toxicity guidelines referenced:
                  TG 414 -- Prenatal Developmental Toxicity Study;
                  TG 443 -- Extended One-Generation Reproductive Toxicity Study}
}

@techreport{OECD_Neurotoxicity,
  author       = {{OECD}},
  title        = {OECD Guidelines for the Testing of Chemicals, Section 4: Health Effects},
  institution  = {Organisation for Economic Co-operation and Development},
  year         = {2004--2007},
  address      = {Paris},
  note         = {Neurotoxicity guidelines referenced:
                  TG 424 -- Neurotoxicity Study in Rodents;
                  TG 426 -- Developmental Neurotoxicity Study}
}

@techreport{OECD_AquaticEcotoxicity,
  author       = {{OECD}},
  title        = {OECD Guidelines for the Testing of Chemicals, Section 2: Effects on Biotic Systems},
  institution  = {Organisation for Economic Co-operation and Development},
  year         = {2004--2019},
  address      = {Paris},
  note         = {Aquatic ecotoxicity guidelines referenced:
                  TG 201 -- Freshwater Alga and Cyanobacteria, Growth Inhibition Test;
                  TG 202 -- Daphnia sp. Acute Immobilisation Test;
                  TG 203 -- Fish, Acute Toxicity Test;
                  TG 210 -- Fish, Early-Life Stage Toxicity Test;
                  TG 211 -- Daphnia magna Reproduction Test}
}

@techreport{OECD_TerrestrialEcotoxicity,
  author       = {{OECD}},
  title        = {OECD Guidelines for the Testing of Chemicals, Section 2: Effects on Biotic Systems},
  institution  = {Organisation for Economic Co-operation and Development},
  year         = {2000--2017},
  address      = {Paris},
  note         = {Terrestrial ecotoxicity guidelines referenced:
                  TG 207 -- Earthworm, Acute Toxicity Tests;
                  TG 208 -- Terrestrial Plant Test: Seedling Emergence and Seedling Growth Test;
                  TG 216 -- Soil Microorganisms: Nitrogen Transformation Test;
                  TG 217 -- Soil Microorganisms: Carbon Transformation Test;
                  TG 222 -- Earthworm Reproduction Test;
                  TG 226 -- Predatory Mite Reproduction Test in Soil;
                  TG 245 -- Honey Bee (Apis mellifera) Larval Toxicity Test, Single Exposure}
}

@techreport{OECD_PersistenceBioaccumulation,
  author       = {{OECD}},
  title        = {OECD Guidelines for the Testing of Chemicals, Section 3: Environmental Fate and Behaviour},
  institution  = {Organisation for Economic Co-operation and Development},
  year         = {1992--2012},
  address      = {Paris},
  note         = {Persistence and bioaccumulation guidelines referenced:
                  TG 301 -- Ready Biodegradability;
                  TG 305 -- Bioaccumulation in Fish: Aqueous and Dietary Exposure;
                  TG 310 -- Ready Biodegradability: CO2 in Sealed Vessels (Headspace Test)}
}

@legislation{EU_CLP,
  author       = {{European Parliament and Council of the European Union}},
  title        = {Regulation (EC) No 1272/2008 on Classification, Labelling and Packaging of Substances and Mixtures (CLP Regulation)},
  year         = {2008},
  address      = {Brussels},
  note         = {Implements the UN GHS in the European Union.}
}

@legislation{EU_REACH_AnnexXIII,
  author       = {{European Parliament and Council of the European Union}},
  title        = {Regulation (EC) No 1907/2006 (REACH), Annex XIII: Criteria for the Identification of Persistent, Bioaccumulative and Toxic Substances, and Very Persistent and Very Bioaccumulative Substances},
  year         = {2006},
  address      = {Brussels},
  note         = {PBT/vPvB assessment criteria as amended by Commission Regulation (EU) No 253/2011.}
}

@legislation{EU2018_1881,
  title        = {Commission Regulation ({EU}) 2018/1881 of 3 December 2018 amending Regulation ({EC}) No 1907/2006 of the {European Parliament} and of the {Council} on the Registration, Evaluation, Authorisation and Restriction of Chemicals ({REACH}) as regards Annexes {I}, {III}, {VI}, {VII}, {VIII}, {IX}, {X}, {XI} and {XII} to address nanoforms of substances},
  author       = {{European Commission}},
  journal      = {Official Journal of the European Union},
  volume       = {L 308},
  pages        = {1--20},
  year         = {2018},
  date         = {2018-12-04},
  url          = {https://eur-lex.europa.eu/eli/reg/2018/1881/oj},
  note         = {Document 32018R1881}
}

@techreport{IARC_Monographs,
  author       = {{International Agency for Research on Cancer}},
  title        = {IARC Monographs on the Identification of Carcinogenic Hazards to Humans},
  institution  = {World Health Organization},
  address      = {Lyon},
  year         = {1972--2024},
  note         = {Classification system: Group 1 (carcinogenic), Group 2A (probably carcinogenic), Group 2B (possibly carcinogenic), Group 3 (not classifiable).}
}

@misc{EU_REACH_2006,
  title        = {Regulation ({EC}) {No} 1907/2006 of the {European Parliament} and of the {Council} of 18 {December} 2006 concerning the {Registration}, {Evaluation}, {Authorisation} and {Restriction} of {Chemicals} ({REACH})},
  author       = {{European Union}},
  year         = {2006},
  note         = {Official Journal of the European Union, L 396, 30.12.2006, p. 1--849},
  url          = {https://eur-lex.europa.eu/legal-content/EN/TXT/?uri=CELEX:02006R1907-20220301},
  number       = {2}
}

@misc{EPA_OPPTS_870,
  title        = {Health Effects Test Guidelines ({OCSPP}/{OPPTS} 870 Series)},
  author       = {{U.S. Environmental Protection Agency}},
  year         = {1998},
  note         = {Office of Chemical Safety and Pollution Prevention ({OCSPP}), formerly Office of Prevention, Pesticides and Toxic Substances ({OPPTS})},
  url          = {https://www.epa.gov/test-guidelines-pesticides-and-toxic-substances/series-870-health-effects-test-guidelines},
  number       = {4}
}

@article{EFSA_MOE_2012,
  title        = {Guidance on the use of the {Margin} of {Exposure} ({MOE}) approach in risk assessment},
  author       = {{European Food Safety Authority (EFSA)}},
  journal      = {EFSA Journal},
  year         = {2012},
  volume       = {10},
  number       = {6},
  pages        = {2571},
  doi          = {10.2903/j.efsa.2012.2571},
  url          = {https://efsa.onlinelibrary.wiley.com/doi/abs/10.2903/j.efsa.2012.2571},
  note         = {Reference [6]}
}

@book{IPCCAR6Physical,
  title={Climate Change 2021: The Physical Science Basis},
  author={Masson-Delmotte, Val{\'e}rie and Zhai, Panmao and Pirani, Anna and Connors, Sarah L and P{\'e}an, Clotilde and Berger, Sophie and Caud, Nada and Chen, Ying and Goldfarb, Leah and Gomis, Melissa I and others},
  year={2021},
  publisher={Cambridge University Press},
  address={Cambridge, United Kingdom and New York, NY, USA},
  note={Contribution of Working Group I to the Sixth Assessment Report of the Intergovernmental Panel on Climate Change},
  doi={10.1017/9781009157896}
}

@techreport{IPCC_WGIII,
  author       = {{Intergovernmental Panel on Climate Change}},
  title        = {Climate Change 2022: Mitigation of Climate Change. Contribution of Working Group III to the Sixth Assessment Report of the Intergovernmental Panel on Climate Change},
  institution  = {Cambridge University Press},
  year         = {2022},
  address      = {Cambridge, UK and New York, NY, USA},
  note         = {Edited by P.R. Shukla, J. Skea, R. Slade, A. Al Khourdajie, R. van Diemen, D. McCollum, M. Pathak, S. Some, P. Vyas, R. Fradera, M. Belkacemi, A. Hasija, G. Lisboa, S. Luz, J. Malley.}
}

@article{huynh2024potential,
  title={The Potential Environmental and Climate Impacts of Stratospheric Aerosol Injection: A Review},
  author={Huynh, Han N and McNeill, V Faye},
  journal={Environmental Science: Atmospheres},
  volume={4},
  number={2},
  pages={114--143},
  year={2024},
  publisher={Royal Society of Chemistry},
  doi={10.1039/D3EA00134B}
}

@report{smith2024state,
  title={The State of Carbon Dioxide Removal},
  author={Smith, Stephen M and Geden, Oliver and Nemet, Gregory F and Gidden, Matthew J and Lamb, William F and Powis, Catherine and Bellamy, Rob and Callaghan, Max W and Cowie, Annette and Cox, Emily and Fuss, Sabine and Gasser, Thomas and Grassi, Giacomo and Greene, Joanna and L{\"u}ck, Stephanie and Moesinger, Alexandra and Müller-Hansen, Finn and Peters, Glen P and Pratama, Yama and Repke, Tim and Riahi, Keywan and Schenuit, Felix and Steinhauser, J{\"o}ran and Strefler, Jessica and Vicu{\~n}a, Jose M and Minx, Jan C},
  year={2024},
  edition={2nd},
  institution={University of Oxford, Global Carbon Project, CO2RE, 4c Carbon},
  url={https://www.stateofcdr.org},
  doi={10.17605/OSF.IO/BFEXG}
}

@article{macmartin2014spatial,
  author  = {MacMartin, Douglas G. and Caldeira, Ken and Keith, David W.},
  title   = {Spatial Scales of Climate Response to Inhomogeneous Radiative Forcing},
  journal = {Journal of Climate},
  year    = {2014},
  volume  = {27},
  number  = {20},
  pages   = {7625--7637},
  doi     = {10.1175/JCLI-D-13-00624.1}
}

@article{fuglestvedt2003metrics,
  author  = {Fuglestvedt, Jan S. and Berntsen, Terje K. and Godal, Odd and Sausen, Robert and Shine, Keith P. and Skodvin, Tora},
  title   = {Metrics of Climate Change: Assessing Radiative Forcing and Emission Indices},
  journal = {Climatic Change},
  year    = {2003},
  volume  = {58},
  number  = {1--2},
  pages   = {267--331},
  doi     = {10.1023/A:1023905326842}
}

@article{blackstock2009climate,
  author        = {Blackstock, Jason J. and others},
  title         = {Climate engineering responses to climate emergencies},
  journal       = {arXiv preprint arXiv:0907.5140},
  year          = {2009},
  eprint        = {0907.5140},
  archivePrefix = {arXiv},
  url           = {https://arxiv.org/abs/0907.5140}
}

@BOOK{2021NASEM.RS,
  author    = {{National Academies of Sciences, Engineering, and Medicine}},
  title     = "{Reflecting Sunlight: Recommendations for Solar Geoengineering Research and Research Governance}",
  year      = 2021,
  publisher = {National Academies Press},
  address   = {Washington, DC, USA},
  doi       = {10.17226/25762},
  url       = {https://nap.nationalacademies.org/catalog/25762/reflecting-sunlight},
  note      = {Committee on Developing a Research Agenda and Research Governance Approaches for Climate Intervention Strategies that Reflect Sunlight to Cool Earth}
}

@REPORT{2023OSTP.SRM,
  author      = {{Office of Science and Technology Policy}},
  title       = "{Congressionally Mandated Research Plan and an Initial Research Governance Framework Related to Solar Radiation Modification}",
  institution = {Office of Science and Technology Policy, Executive Office of the President},
  address     = {Washington, DC, USA},
  year        = 2023,
  month       = jun,
  url         = {https://bidenwhitehouse.archives.gov/wp-content/uploads/2023/06/Congressionally-Mandated-Report-on-Solar-Radiation-Modification.pdf},
  note        = {Accessed 15 September 2025}
}

@article{eastham2025key,
  author    = {Eastham, Sebastian D. and others},
  title     = {Key gaps in models' physical representation of climate intervention and its impacts},
  journal   = {Journal of Advances in Modeling Earth Systems},
  volume    = {17},
  number    = {6},
  pages     = {e2024MS004872},
  year      = {2025},
  publisher = {American Geophysical Union},
  doi       = {10.1029/2024MS004872}
}

@article{Pongratz2012,
  author  = {Pongratz, Julia and Lobell, David B. and Cao, Long and Caldeira, Ken},
  title   = {Crop yields in a geoengineered climate},
  journal = {Nature Climate Change},
  volume  = {2},
  pages   = {101--105},
  year    = {2012},
  doi     = {10.1038/nclimate1373}
}

@article{Loeb2018,
  author  = {Loeb, Norman G. and Doelling, David R. and Wang, Hailan and Su, Wenying and Nguyen, Chinh and Corbett, Joseph G. and Liang, Lusheng and Mitrescu, Cristian and Rose, Fred G. and Kato, Seiji},
  title   = {Clouds and the {Earth's} Radiant Energy System ({CERES}) Energy Balanced and Filled ({EBAF}) Top-of-Atmosphere ({TOA}) Edition-4.0 Data Product},
  journal = {Journal of Climate},
  volume  = {31},
  number  = {2},
  pages   = {895--918},
  year    = {2018},
  doi     = {10.1175/JCLI-D-17-0208.1}
}

@article{Myhre2025,
  author  = {Myhre, Gunnar and Hodnebrog, {\O}ivind and Loeb, Norman and Forster, Piers M.},
  title   = {Observed trend in {Earth} energy imbalance may provide a constraint for low climate sensitivity models},
  journal = {Science},
  volume  = {388},
  number  = {6752},
  pages   = {1210--1213},
  year    = {2025},
  doi     = {10.1126/science.adt0647}
}

@article{Stephens2012,
  author  = {Stephens, Graeme L. and Li, Juilin and Wild, Martin and Clayson, Carol Anne and Loeb, Norman and Kato, Seiji and L'Ecuyer, Tristan and Stackhouse, Paul W. and Lebsock, Matthew and Andrews, Timothy},
  title   = {An update on {Earth's} energy balance in light of the latest global observations},
  journal = {Nature Geoscience},
  volume  = {5},
  number  = {10},
  pages   = {691--696},
  year    = {2012},
  doi     = {10.1038/ngeo1580}
}

@REPORT{2023UNEP.SRMReport,
  author      = {{United Nations Environment Programme}},
  title       = "{One Atmosphere: An Independent Expert Review on Solar Radiation Modification Research and Deployment}",
  institution = {United Nations Environment Programme},
  address     = {Nairobi, Kenya},
  year        = 2023,
  month       = feb,
  day         = 28,
  url         = {https://www.unep.org/resources/report/Solar-Radiation-Modification-research-deployment},
  note        = {Accessed 15 September 2025}
}

@REPORT{2025RoyalSoc.SRM,
  author      = {{The Royal Society}},
  title       = "{Solar Radiation Modification: Policy Briefing}",
  institution = {The Royal Society},
  address     = {London, UK},
  year        = 2025,
  url         = {https://royalsociety.org/news-resources/projects/solar-radiation-modification/},
  note        = {Accessed 15 September 2025; independent scientific policy briefing on SRM techniques and implications}  
}

@article{dykema2016improved,
  title={Improved aerosol radiative properties as a foundation for solar geoengineering risk assessment},
  author={Dykema, John A and Keith, David W and Keutsch, Frank N},
  journal={Geophysical Research Letters},
  volume={43},
  number={14},
  pages={7758--7766},
  year={2016},
  doi={10.1002/2016GL069258},
  publisher={Wiley Online Library}
}

@inproceedings{teller1997global,
  title={Global Warming and Ice Ages: I. Prospects for Physics-Based Modulation of Global Change},
  author={Teller, Edward and Wood, Lowell and Hyde, Roderick},
  booktitle={International Symposium on Planetary Emergencies},
  address={Erice, Italy},
  year={1997}
}

@article{crutzen2006albedo,
  author    = {Crutzen, Paul J.},
  title     = {Albedo Enhancement by Stratospheric Sulfur Injections: A Contribution to Resolve a Policy Dilemma?},
  journal   = {Climatic Change},
  volume    = {77},
  number    = {3--4},
  pages     = {211--220},
  year      = {2006},
  doi       = {10.1007/s10584-006-9101-y}
}

@article{pope2012stratospheric,
  title={Stratospheric aerosol particles and solar-radiation management},
  author={Pope, Francis D and Braesicke, Peter and Grainger, Roy G and Kalberer, Markus and Watson, I Matthew and Davidson, Peter J and Cox, Richard A},
  journal={Nature Climate Change},
  volume={2},
  number={10},
  pages={713--719},
  year={2012},
  doi = {10.1038/nclimate1528}, 
  publisher={Nature Publishing Group}
}

@article{colarco2010online,
  title={Online simulations of global aerosol distributions in the NASA GEOS-4 model and comparisons to satellite and ground-based aerosol optical depth},
  author={Colarco, Peter and da Silva, Arlindo and Chin, Mian and Diehl, Thomas},
  journal={Journal of Geophysical Research: Atmospheres},
  volume={115},
  number={D14},
  year={2010},
  publisher={Wiley Online Library}
}

@techreport{sparc2006assessment,
  title={SPARC Assessment of Stratospheric Aerosol Properties (ASAP)},
  author={{SPARC}},
  editor={Thomason, L and Peter, Th},
  institution={SPARC},
  number={Report No. 4, WCRP-124, WMO/TD-No. 1295},
  year={2006}
}

@article{weisenstein2015solar,
  title={Solar geoengineering using solid aerosol in the stratosphere},
  author={Weisenstein, Debra K and Keith, David W and Dykema, John A},
  journal={Atmospheric Chemistry and Physics},
  volume={15},
  number={20},
  pages={11835--11859},
  year={2015},
  publisher={Copernicus GmbH}
}

@article{rasch2008overview,
  title={An overview of geoengineering of climate using stratospheric sulphate aerosols},
  author={Rasch, Philip J and Tilmes, Simone and Turco, Richard P and Robock, Alan and Oman, Luke and Chen, Chih-Chieh and Stenchikov, Georgiy L and Garcia, Rolando R},
  journal={Philosophical Transactions of the Royal Society A: Mathematical, Physical and Engineering Sciences},
  volume={366},
  number={1882},
  pages={4007--4037},
  doi = {10.1098/rsta.2008.0131}, 
  year={2008},
  publisher={The Royal Society London}
}

@techreport{sparc2017data,
  title={The SPARC Data Initiative: Assessment of stratospheric trace gas and aerosol climatologies from satellite limb sounders},
  author={{SPARC}},
  editor={Hegglin, Michaela I and Tegtmeier, Susann},
  institution={SPARC},
  number={Report No. 8, WCRP-5/2017},
  year={2017}
}

@article{tilmes2024research,
  title={Research criteria towards an interdisciplinary Stratospheric Aerosol Intervention assessment},
  author={Tilmes, Simone and Rosenlof, Karen H and Visioni, Daniele and Bednarz, Ewa M and others},
  journal={Oxford Open Climate Change},
  volume={4},
  number={1},
  pages={kgae010},
  year={2024},
  doi={10.1093/oxfclm/kgae010},
  publisher={Oxford University Press}
}

@article{aquila2014modifications,
  title={Modifications of the quasi-biennial oscillation by a geoengineering perturbation of the stratospheric aerosol layer},
  author={Aquila, Valentina and Garfinkel, Chaim I and Newman, Paul A and Oman, Luke D and Waugh, Darryn W},
  journal={Geophysical Research Letters},
  volume={41},
  number={5},
  pages={1738--1744},
  doi = {10.1002/2013GL058818},
  year={2014},
  publisher={Wiley Online Library}
}

@article{Jones2022,
  author    = {Jones, Andy and Haywood, Jim M. and Scaife, Adam A. and Boucher, Olivier and Henry, Matthew and Kravitz, Ben and Lurton, Thibaut and Nabat, Pierre and Niemeier, Ulrike and S{\'e}f{\'e}rian, Roland and Tilmes, Simone and Visioni, Daniele},
  title     = {The impact of stratospheric aerosol intervention on the {N}orth {A}tlantic and {Q}uasi-{B}iennial {O}scillations in the {G}eoengineering {M}odel {I}ntercomparison {P}roject ({GeoMIP}) {G6sulfur} experiment},
  journal   = {Atmospheric Chemistry and Physics},
  volume    = {22},
  pages     = {2999--3016},
  year      = {2022},
  doi       = {10.5194/acp-22-2999-2022},
  url       = {https://acp.copernicus.org/articles/22/2999/2022/}
}

@article{haywood2025wcrp,
  title={World Climate Research Programme lighthouse activity: an assessment of major research gaps in solar radiation modification research},
  author={Haywood, Jim M and others},
  journal={Frontiers in Climate},
  volume={7},
  pages={1507479},
  year={2025},
  doi = {10.3389/fclim.2025.1507479},
  publisher={Frontiers}
}

@article{bala2008impact,
  title={Impact of geoengineering schemes on the global hydrological cycle},
  author={Bala, Govindasamy and Duffy, Philip B and Taylor, Karl E},
  journal={Proceedings of the National Academy of Sciences},
  volume={105},
  number={22},
  pages={7664--7669},
  year={2008},
  doi = {10.1073/pnas.0711648105},
  publisher={National Acad Sciences}
}

@article{Free2009,
  author  = {Free, Melissa and Lanzante, John R.},
  title   = {Effect of Volcanic Eruptions on the Vertical Temperature Profile in Radiosonde Data and Climate Models},
  journal = {Journal of Climate},
  year    = {2009},
  volume  = {22},
  number  = {11},
  pages   = {2925--2939},
  doi     = {10.1175/2008JCLI2562.1}
}

@article{Nowack2023,
  author    = {Nowack, Peer and Ceppi, Paulo and Davis, Sean M. and Chiodo, Gabriel and Ball, Will and Diallo, Mohamadou A. and Hassler, Birgit and Jia, Yue and Keeble, James and Joshi, Manoj},
  title     = {Response of stratospheric water vapour to warming constrained by satellite observations},
  journal   = {Nature Geoscience},
  volume    = {16},
  pages     = {577--583},
  year      = {2023},
  doi       = {10.1038/s41561-023-01183-6}
}

@article{Cheng2022,
  author  = {Cheng, Wei and MacMartin, Douglas G. and Kravitz, Ben and Visioni, Daniele and Bednarz, Ewa M. and Xu, Yangyang and Luo, Yong and Huang, Lei and Hu, Yongyun and Staten, Paul W. and Hitchcock, Peter and Moore, John C. and Guo, Anboyu and Deng, Xiangzheng},
  title   = {Changes in {Hadley} circulation and intertropical convergence zone under strategic stratospheric aerosol geoengineering},
  journal = {npj Climate and Atmospheric Science},
  year    = {2022},
  volume  = {5},
  pages   = {32},
  doi     = {10.1038/s41612-022-00254-6}
}

@article{Simpson2019,
  author  = {Simpson, Isla R. and Tilmes, Simone and Richter, Jadwiga H. and Kravitz, Ben and MacMartin, Douglas G. and Mills, Michael J. and Fasullo, John T. and Pendergrass, Angeline G.},
  title   = {The Regional Hydroclimate Response to Stratospheric Sulfate Geoengineering and the Role of Stratospheric Heating},
  journal = {Journal of Geophysical Research: Atmospheres},
  year    = {2019},
  volume  = {124},
  number  = {23},
  pages   = {12587--12616},
  doi     = {10.1029/2019JD031093}
}

@article{Loeb2025,
  author  = {Loeb, Norman G. and Thorsen, Tyler J. and Kato, Seiji and Rose, Fred G. and Hodnebrog, {\O}ivind and Myhre, Gunnar},
  title   = {Emerging hemispheric asymmetry of {Earth}'s radiation},
  journal = {Proceedings of the National Academy of Sciences},
  year    = {2025},
  volume  = {122},
  number  = {40},
  pages   = {e2511595122},
  doi     = {10.1073/pnas.2511595122}
}

@article{Roose2023,
  author  = {Roose, Shinto and Bala, Govindasamy and Krishnamohan, K. S. and Cao, Long and Caldeira, Ken},
  title   = {Quantification of tropical monsoon precipitation changes in terms of interhemispheric differences in stratospheric sulfate aerosol optical depth},
  journal = {Climate Dynamics},
  year    = {2023},
  volume  = {61},
  pages   = {4243--4258},
  doi     = {10.1007/s00382-023-06799-3}
}

@article{Visioni2023GeoMIP,
  author  = {Visioni, Daniele and Kravitz, Ben and Robock, Alan and Tilmes, Simone and Haywood, Jim and Boucher, Olivier and Lawrence, Mark and Irvine, Peter and Niemeier, Ulrike and Xia, Lili and Chiodo, Gabriel and Lennard, Chris and Watanabe, Shingo and Moore, John C. and Muri, Helene},
  title   = {Opinion: The scientific and community-building roles of the {G}eoengineering {M}odel {I}ntercomparison {P}roject ({GeoMIP}) -- past, present, and future},
  journal = {Atmospheric Chemistry and Physics},
  year    = {2023},
  volume  = {23},
  pages   = {5149--5176},
  doi     = {10.5194/acp-23-5149-2023}
}

@article{Zhang2024hemispheric,
  author  = {Zhang, Yan and MacMartin, Douglas G. and Visioni, Daniele and Bednarz, Ewa M. and Kravitz, Ben},
  title   = {Hemispherically symmetric strategies for stratospheric aerosol injection},
  journal = {Earth System Dynamics},
  year    = {2024},
  volume  = {15},
  pages   = {191--213},
  doi     = {10.5194/esd-15-191-2024}
}

@article{Tracy2022health,
  author  = {Tracy, Sarah M. and Moch, Jonathan M. and Eastham, Sebastian D. and Buonocore, Jonathan J.},
  title   = {Stratospheric aerosol injection may impact global systems and human health outcomes},
  journal = {Elementa: Science of the Anthropocene},
  year    = {2022},
  volume  = {10},
  number  = {1},
  pages   = {00047},
  doi     = {10.1525/elementa.2022.00047}
}

@article{Proctor2018crops,
  author  = {Proctor, Jonathan and Hsiang, Solomon and Burney, Jennifer and Burke, Marshall and Schlenker, Wolfram},
  title   = {Estimating global agricultural effects of geoengineering using volcanic eruptions},
  journal = {Nature},
  year    = {2018},
  volume  = {560},
  pages   = {480--483},
  doi     = {10.1038/s41586-018-0417-3}
}

@article{Bednarz2023ozone,
  author  = {Bednarz, Ewa M. and Butler, Amy H. and Visioni, Daniele and Zhang, Yan and Kravitz, Ben and MacMartin, Douglas G.},
  title   = {Injection strategy -- a driver of atmospheric circulation and ozone response to stratospheric aerosol geoengineering},
  journal = {Atmospheric Chemistry and Physics},
  year    = {2023},
  volume  = {23},
  pages   = {13665--13684},
  doi     = {10.5194/acp-23-13665-2023}
}

@techreport{ECHA_R8,
  author       = {{ECHA}},
  title        = {Guidance on Information Requirements and Chemical Safety 
                  Assessment. Chapter {R.8}: Characterisation of Dose 
                  [Concentration]--Response for Human Health},
  institution  = {European Chemicals Agency},
  address      = {Helsinki, Finland},
  year         = {2012},
  url          = {https://echa.europa.eu/guidance-documents/guidance-on-information-requirements-and-chemical-safety-assessment}
}

@article{Coddington2019,
  author  = {Coddington, O. and Lean, J. and Pilewskie, P. and Snow, M. and Lindholm, D.},
  title   = {Solar Irradiance Variability: Comparisons of Models and Measurements},
  journal = {Earth and Space Science},
  volume  = {6},
  pages   = {2525--2555},
  year    = {2019},
  doi     = {10.1029/2019EA000693}
}

@techreport{NOAA2002_OzoneAssessment,
  author       = {{National Oceanic and Atmospheric Administration}},
  title        = {Scientific Assessment of Ozone Depletion: 2002},
  institution  = {World Meteorological Organization and United Nations Environment Programme},
  year         = {2003},
  number       = {47},
  address      = {Geneva, Switzerland},
  note         = {Global Ozone Research and Monitoring Project--Report No. 47},
}

@article{Wells2024strategies,
  author  = {Wells, Annabel F. and Haywood, Jim M. and Bednarz, Ewa M. and Visioni, Daniele and MacMartin, Douglas G.},
  title   = {Identifying Climate Impacts From Different Stratospheric Aerosol Injection Strategies in {UKESM1}},
  journal = {Earth's Future},
  year    = {2024},
  volume  = {12},
  pages   = {e2023EF004358},
  doi     = {10.1029/2023EF004358}
}

@article{Franke2021QBO,
  author  = {Franke, Henning and Niemeier, Ulrike and Visioni, Daniele},
  title   = {Differences in the quasi-biennial oscillation response to stratospheric aerosol modification depending on injection strategy and species},
  journal = {Atmospheric Chemistry and Physics},
  year    = {2021},
  volume  = {21},
  pages   = {8615--8638},
  doi     = {10.5194/acp-21-8615-2021}
}

@article{Keys2022perceivedfailure,
  author  = {Keys, Patrick W. and Barnes, Elizabeth A. and Diffenbaugh, Noah S. and Hurrell, James W. and Bell, Claudia M.},
  title   = {Potential for perceived failure of stratospheric aerosol injection deployment},
  journal = {Proceedings of the National Academy of Sciences},
  year    = {2022},
  volume  = {119},
  number  = {40},
  pages   = {e2210036119},
  doi     = {10.1073/pnas.2210036119}
}

@article{Kovilakam2020GloSSAC,
  author  = {Kovilakam, Mahesh and Thomason, Larry W. and Ernest, Natalie and Rieger, Landon and Bourassa, Adam and Mill{\'a}n, Luis},
  title   = {The {G}lobal {S}pace-based {S}tratospheric {A}erosol {C}limatology (version 2.0): 1979--2018},
  journal = {Earth System Science Data},
  year    = {2020},
  volume  = {12},
  pages   = {2607--2634},
  doi     = {10.5194/essd-12-2607-2020}
}

@article{Kovilakam2023SAGEIII,
  author  = {Kovilakam, Mahesh and Thomason, Larry W. and Knepp, Travis},
  title   = {{SAGE III/ISS} aerosol/cloud categorization and its impact on {GloSSAC}},
  journal = {Atmospheric Measurement Techniques},
  year    = {2023},
  volume  = {16},
  pages   = {2709--2731},
  doi     = {10.5194/amt-16-2709-2023}
}

@article{Lo2016detection,
  author  = {Lo, Y. T. Eunice and Charlton-Perez, Andrew J. and Lott, Fraser C. and Highwood, Eleanor J.},
  title   = {Detecting sulphate aerosol geoengineering with different methods},
  journal = {Scientific Reports},
  year    = {2016},
  volume  = {6},
  pages   = {39169},
  doi     = {10.1038/srep39169}
}

@article{Kravitz2020uncertainty,
  author  = {Kravitz, Ben and MacMartin, Douglas G.},
  title   = {Uncertainty and the basis for confidence in solar geoengineering research},
  journal = {Nature Reviews Earth \& Environment},
  year    = {2020},
  volume  = {1},
  number  = {1},
  pages   = {64--75},
  doi     = {10.1038/s43017-019-0004-7}
}

@techreport{WMO2022,
  author      = {{World Meteorological Organization}},
  title       = {Scientific Assessment of Ozone Depletion: 2022},
  institution = {WMO},
  year        = {2022},
  type        = {GAW Report},
  number      = {278},
  address     = {Geneva, Switzerland},
  pages       = {509},
}

@article{Dietmuller2021,
  author  = {Dietm{\"u}ller, S. and Garny, H. and Eichinger, R. 
             and Ball, W. T.},
  title   = {Analysis of recent lower-stratospheric ozone trends in 
             chemistry climate models},
  journal = {Atmospheric Chemistry and Physics},
  volume  = {21},
  pages   = {6811--6837},
  year    = {2021},
  doi     = {10.5194/acp-21-6811-2021},
}

@article{BenitoBarca2025,
  author  = {Benito-Barca, S. and Abalos, M. and Calvo, N.},
  title   = {Recent Lower Stratospheric Ozone Trends in {CCMI-2022} 
             Models: Role of Natural Variability and Transport},
  journal = {Journal of Geophysical Research: Atmospheres},
  volume  = {130},
  pages   = {e2024JD042412},
  year    = {2025},
  doi     = {10.1029/2024JD042412},
}

@article{Abalos2026,
  author  = {Abalos, M. and Birner, T. and Chrysanthou, A. and Davis, S. 
             and de la C{\'a}mara, A. and Dhomse, S. and Garny, H. 
             and Hegglin, M. I. and Hubert, D. and Ivaniha, O. 
             and Keeble, J. and Linz, M. and Minganti, D. and Neu, J. 
             and Plummer, D. and Saunders, L. and Shah, K. and Stiller, G. 
             and Tourpali, K. and Waugh, D. and others},
  title   = {Evaluation of stratospheric transport in three generations 
             of {Chemistry--Climate Models}},
  journal = {EGUsphere [preprint]},
  year    = {2026},
  doi     = {10.5194/egusphere-2025-6549},
  note    = {In review for Atmospheric Chemistry and Physics},
}

@article{Tilmes2022,
  author  = {Tilmes, S. and Visioni, D. and Jones, A. and Haywood, J. 
             and S{\'e}f{\'e}rian, R. and Nabat, P. and Boucher, O. 
             and Bednarz, E. M. and Niemeier, U.},
  title   = {Stratospheric ozone response to sulfate aerosol and solar 
             dimming climate interventions based on the {G6} 
             {Geoengineering Model Intercomparison Project} ({GeoMIP}) 
             simulations},
  journal = {Atmospheric Chemistry and Physics},
  volume  = {22},
  pages   = {4557--4579},
  year    = {2022},
  doi     = {10.5194/acp-22-4557-2022},
}

@article{Vattioni2025,
  author  = {Vattioni, S. and Peter, T. and Weber, R. and Dykema, J. A. 
             and Luo, B. and Stenke, A. and Feinberg, A. 
             and Sukhodolov, T. and Keutsch, F. N. and Ammann, M. 
             and Vockenhuber, C. and D{\"o}beli, M. 
             and Kelesidis, G. A. and Chiodo, G.},
  title   = {Injecting solid particles into the stratosphere could 
             mitigate global warming but currently entails great 
             uncertainties},
  journal = {Communications Earth \& Environment},
  volume  = {6},
  pages   = {132},
  year    = {2025},
  doi     = {10.1038/s43247-025-02038-1},
}
\end{document}